\documentclass[%
 reprint,
showpacs,preprintnumbers,
 amsmath,amssymb,
 aps,
]{revtex4-1}

\usepackage{graphicx}

\newcommand{\myfig}[4][ht]{
\begin{figure}[#1]
\centering
\includegraphics[#2]{#3}
\caption{#4\label{#3}}
\end{figure}
}

\newcommand{\myfigwide}[4][ht]{
\begin{figure*}[#1]
\centering
\includegraphics[#2]{#3}
\caption{#4\label{#3}}
\end{figure*}
}

\newcommand{\npar}{%
\\[-3mm]\par}

\newcommand{\revision}[1]{%
{#1}%
}

\graphicspath{{images/}}

\usepackage{dcolumn}
\usepackage{bm}

\begin{document}


\title{SUPERDIFFUSIVE HEAT CONDUCTION IN SEMICONDUCTOR ALLOYS\\[2mm]I. Theoretical foundations}


\author{Bjorn Vermeersch$^1$}
\email{bvermeer@purdue.edu}
\author{Jes\'{u}s Carrete$^2$}
\author{Natalio Mingo$^2$}
\author{Ali Shakouri$^1$}
\email{shakouri@purdue.edu}
\affiliation{\vspace{3mm}$^1$ Birck Nanotechnology Center, Purdue University, West Lafayette, Indiana 47907, USA\\
$^2$ LITEN, CEA-Grenoble, 38054 Grenoble Cedex 9, France}

\date{\today}

\begin{abstract}
Semiconductor alloys exhibit a strong dependence of effective thermal conductivity on measurement frequency. So far this quasi-ballistic behaviour has only been interpreted phenomenologically, providing limited insight into the underlying thermal transport dynamics. Here, we show that quasi-ballistic heat conduction in semiconductor alloys is governed by L\'evy superdiffusion. By solving the Boltzmann transport equation (BTE) with ab initio phonon dispersions and scattering rates, we reveal a transport regime \revision{with fractal space dimension $1 < \alpha < 2$ and superlinear time evolution} of mean square energy displacement $\sigma^2(t) \sim t^{\beta} (1 < \beta < 2)$. The \revision{characteristic} exponent\revision{s are} directly interconnected with the order $n$ of the dominant phonon scattering mechanism $\tau \sim \omega^{-n} (n>3)$ and \revision{cumulative conductivity spectra} $\kappa_{\Sigma}(\tau;\Lambda)\sim (\tau;\Lambda)^{\gamma}$ \revision{resolved for relaxation times or mean free paths} through simple relations $\alpha = 3-\beta = 1 + 3/n = 2 - \gamma$. The quasi-ballistic transport inside alloys is no longer governed by Brownian motion, but instead \revision{dominated by} L\'evy \revision{dynamics}. This has important implications for the interpretation of thermoreflectance (TR) measurements with modified Fourier theory. Experimental $\alpha$ values for InGaAs and SiGe, determined through TR analysis with a novel L\'evy heat formalism, match ab initio BTE predictions within a few percent. Our findings lead to a deeper and more accurate quantitative understanding of the physics of nanoscale heat flow experiments.
\end{abstract}
\pacs{65.40.-b , 63.20.-e}%
\maketitle


\subsection*{Introduction}
\revision{Heat conduction in solid media is normally described by the well known Fourier diffusion equation. However, at length and time scales comparable respectively to phonon mean free paths (MFPs) $\Lambda$ and relaxation times $\tau$,} this classical model begins to fail \cite{boltzmann}. Experimental observations in which thermal gradients are induced over microscale distances clearly deviate from standard Fourier predictions \cite{cahill}\nocite{siemens}\nocite{minnich}\nocite{johnson}--\cite{malen}. A notable and pioneering example consists of time domain thermoreflectance (TDTR) measurements on semiconductor alloys \cite{cahill}. This technique performs thermal characterisation of test samples by measuring their response to a modulated pulse train of ultrashort laser pulses. Alloys including InGaAs and SiGe were found to exhibit a near 50\% reduction in apparent Fourier thermal conductivity over the 1--10$\,$MHz laser modulation frequency range. Such effects, and similar ones observed in other experimental configurations, can be attributed to phonons whose MFPs exceed the characteristic dimension of the heat source \cite{cahill}--\cite{collins}, in this case the frequency dependent thermal penetration depth. As these so called quasi-ballistic modes are unlikely to experience scattering events within the thermal gradient, they violate the inherent assumption of the Fourier model, causing it to fail. \revision{Recently, Wilson and coworkers proposed a phenomenological two-channel model \cite{wilson} to explain the nondiffusive behaviour in semiconductor alloys. This approach divides the phonon population into two parts (essentially long and short MFP modes) which are treated as individual conduction channels with distinct temperatures. The channel equilibration over space and time is governed by their individual thermal conductivities and energy exchange coupling factor, all of which serve as fitting parameters in the interpretation of experimental data.}
\npar%
In this work, we \revision{demonstrate that such an artificial division of the phonon population and underlying assumption of internal nonequilibrium are unnecessary. Rather, the Fourier breakdown in alloys can be easily understood in terms of an anomalous single temperature regime that arises naturally due to the inherent properties of mass impurity phonon scattering.} We use analytical solutions of the Boltzmann transport equation (BTE) under the relaxation time approximation to study the \revision{spatiotemporal dynamics of quasi-ballistic heat conduction} in close detail. Our results reveal that the earlier mentioned reductions of apparent conductivity in semiconductor alloys are a direct manifestation of superdiffusive \revision{L\'evy transport. In this regime, the thermal field is characterised by a fractal space dimension $1 < \alpha < 2$ that can be experimentally measured and a mean square energy displacement that grows with superlinear time exponent $\beta = 3-\alpha$. The value of the exponents} as well as the extent of the regime \revision{in space and modulation frequency} are closely related to the shape of the cumulative conductivity $\kappa_{\Sigma}(\Lambda^{\ast})$ and $\kappa_{\Sigma}(\tau^{\ast})$. These functions are the conductivity that would be observed if only those phonon modes with \revision{MFPs} $\Lambda \leq \Lambda^{\ast}$ \revision{or} scattering times $\tau \leq \tau^{\ast}$ were present in the medium. The \revision{L\'evy} energy density is no longer Gaussian, causing shortcomings of modified Fourier models conventionally used for experimental analysis of quasi-ballistic heat flow.
\subsection*{BTE modelling}
To identify key regimes of thermal transport inside the semiconductor as would occur during thermoreflectance experiments, we solve the 1D BTE for cross-plane heat flow ($z$ direction) \revision{under the relaxation time approximation (RTA). The RTA hypothesis treats all phonon scattering processes as resistive, which has been shown to be an acceptable simplification (with an error typically below 5--10\%) for Si and III-V semiconductor compounds \cite{ShengBTE_2014}. As is customary \cite{phonons}, we assume the system close to thermal equilibrium such that a meaningful local temperature rise $\Delta T(z,t) = T(z,t) - T_0$ can be defined with respect to the reference temperature $T_0$. All experimental thermoreflectance settings take care to ensure that $\Delta T \ll T_0$. The deviational volumetric energy density $\hbar \omega_k [f_{\text{BE}}(\omega_k;T) - f_{\text{BE}}(\omega_k;T_0)]/V$ for a phonon mode with wavevector $\vec{k}$ and frequency $\omega$ can then be linearised to $C_k \Delta T$, where $C_k = (\hbar \omega_k/V) \frac{\partial f_{\text{BE}}}{\partial T}\bigr|_{T_0}$ denotes the mode specific heat with $V$ the volume of the supercell and $f_{\text{BE}}$ the Bose-Einstein distribution. Under these assumptions,} the single pulse response to a planar source located at $z=0$ with unit strength is described by:
\begin{equation}
\frac{\partial g_k}{\partial t} + v_{\shortparallel,k} \frac{\partial g_k}{\partial z} = - \frac{g_k - C_k \Delta T}{\tau_k} + \frac{C_k}{C} \delta(z) \delta(t) \label{BTEmaster}
\end{equation}
$g_k$ denotes the deviational phonon energy distribution and $v_{\shortparallel,k} = \frac{\partial \omega}{\partial k_z}$ is the projection of the phonon group velocity onto the thermal transport axis. Note that we write the BTE in terms of spectral quantities, indicated by subscripts $_k$, resolved for phonon wavevector $\vec{k}$ (and, implicitly, phonon branch). \revision{This is in contrast to other recent works \cite{minnichBTE,collins} that rely on phonon frequency $\omega$ as the main variable under the assumption of a spherically symmetric Brillouin zone (BZ).} Our approach enables to directly utilise full phonon dispersions obtained from ab initio calculations and thereby accounts for the actual BZ shape and potential crystal anisotropies. The scaling factor $(C_k/C)$ in the source term expresses that the injected energy gets distributed across the various phonon modes in proportionality with their contribution to the total heat capacity $C = \sum \limits_k C_k$. After spatial Fourier ($z \leftrightarrow \xi$) and temporal Laplace ($t \leftrightarrow s$) transformation of Eq. (\ref{BTEmaster}) we obtain
\begin{equation}
G_k(\xi,s) = \frac{C_k \left[ \Delta T(\xi,s) + \frac{\tau_k}{C} \right]}{1 + s \tau_k + j \xi \Lambda_{\shortparallel,k}} \label{GkRTA}
\end{equation}
where $j$ is the complex unit and $\Lambda_{\shortparallel,k} = v_{\shortparallel,k} \cdot \tau_k$ denotes the projected phonon MFP. \revision{We note that the usefulness of $G_k$ is not limited to the RTA. Analysis of the full BTE leads to a large system of linear equations in ${G_k}$ which can be solved efficiently by iterative refinement \cite{omini_beyond_1996,li_thermal_2012} of the RTA solution (\ref{GkRTA}). This may be explored further in future work.} By inserting $G_k$ into the conservation of energy,
\begin{equation}
\sum \limits_k \frac{1}{\tau_k} \left( G_k - C_k \Delta T \right) = 0 \quad ,
\end{equation}
we can solve for $\Delta T(\xi,s)$. In the above and following, we use summations $\sum \limits_k$ of all branches over a discrete wavevector grid as this is convenient for ab initio simulation data. Presented BTE expressions are readily applicable to analytical modelling as well by simply exchanging the summations with volume integrals $\iiint \limits_{\text{BZ}} \mathrm{d}\vec{k}$. Due to crystal symmetry with respect to the $k_z=0$ plane, projected phonon MFPs occur in pairs with identical magnitude but opposite sign, and we obtain:
\begin{equation}
P(\xi,s) = C \Delta T(\xi,s) = \frac{\sum \limits_{k_z \geq 0} C_k \Psi_k(\xi,s)}{\sum \limits_{k_z \geq 0} \frac{C_k}{\tau_k} \left[ 1 - \Psi_k(\xi,s) \right]} \label{Pmaster}
\end{equation}
in which
\begin{equation}
\Psi_k(\xi,s) = \frac{1 + s \tau_k}{(1 + s \tau_k)^2 + \xi^2 \Lambda_{\shortparallel,k}^2}
\end{equation}
Equation (\ref{Pmaster}) provides a closed form expression in the Fourier-Laplace domain for the 1D energy density single pulse response in the semiconductor. One can easily verify fulfillment of the energy balance by observing that $P(\xi \rightarrow 0,s) = 1/s$, signaling that $\int \limits_{-\infty}^{\infty} P(z,t) \mathrm{d}z = 1$ at all times $t$. In stochastic terms, $P(\xi,s)$ is the characteristic function of a properly normalised random walk process in which $P(z,t)\mathrm{d}z$ expresses the probability to find the injected source energy in location range $[z,z+\mathrm{d}z]$ at time $t$. The derivatives of $P(\xi,s)$ with respect to $\xi$ are continuous at $\xi=0$, so the moments of $P(z,t)$ exist and are finite. The vanishing of the first derivative $\frac{\partial P(\xi,s)}{\partial \xi}\big|_{\xi=0} = 0$ indicates zero mean, in accordance with spatial symmetry of the energy density around the heat source. The mean square displacement (MSD) of thermal energy now immediately follows
\begin{equation}
\sigma^2(s) = - \frac{\partial^2 P(\xi,s)}{\partial \xi^2}\biggr|_{\xi=0} = \frac{2}{s^2} \cdot \frac{\sum \limits_k \frac{\kappa_k}{(1+s\tau_k)^2}}{\sum \limits_k \frac{C_k}{1+s\tau_k}} \label{variancemaster}
\end{equation}
in which we introduced the spectral thermal conductivity $\kappa_k = \Lambda_{\shortparallel,k} \cdot v_{\shortparallel,k} \cdot C_k$. \revision{Physically, the square root of the MSD provides a measure for the average spatial extent of the thermal field. In Fourier diffusion regime, this quantity is commonly known as the thermal penetration length.}
\npar%
From Eq. (\ref{variancemaster}) we readily recover two well established limit regimes of the thermal transport. When $s \tau_{\omega} \gg 1$, corresponding to transitions that occur quickly compared to the phonon relaxation times, we have
\begin{equation}
\sigma^2(s \rightarrow \infty) = \frac{2}{s^3} \bar{v}^2 \Rightarrow \sigma^2(t) = \bar{v}^2 t^2
\end{equation}
in which
\begin{equation}
\bar{v}^2 = \frac{\sum \limits_k \frac{C_k}{\tau_k} v_{\shortparallel,k}^2}{\sum \limits_k \frac{C_k}{\tau_k}} \label{vbar}
\end{equation}
This can be interpreted as purely ballistic energy displacement at net ensemble group velocity $\bar{v}$. At long time scales, $s \tau_{\omega} \ll 1$, on the other hand, 
\begin{equation}
\sigma^2(s \rightarrow 0) = \frac{2}{s^2} \frac{\sum \limits_k \kappa_k}{\sum \limits_k C_k} \Rightarrow \sigma^2(t) = 2 \frac{\kappa}{C} t = 2Dt
\end{equation}
which is the standard diffusive regime. The remainder of the paper studies the transition between ballistic and diffusive limits in detail. We first outline our simulation procedures for obtaining the phonon properties which then serve as input for (\ref{Pmaster}) and (\ref{variancemaster}). We demonstrate the onset of fractal L\'evy transport in semiconductor alloys, explain its physical origin, and point out significant differences between the associated dynamics and modified Fourier solutions.
\subsection*{Ab initio methodology}
We start by considering the fully ordered semiconductors Si, Ge, InAs and GaAs. For each of them, we perform an unconstrained relaxation of their unit cell using the VASP DFT package \cite{vasp} with projector-augmented-wave (PAW) pseudopotentials \cite{PAW}, the local density approximation to exchange and correlation \cite{LDA}, a $12\times 12\times 12$ $\vec{k}$-point grid, and a plane-wave energy cutoff 30\% higher than the maximum value prescibed for each pseudopotential. We then generate a minimal set of displaced $6\times 6\times 6$ supercells using Phonopy \cite{phonopy}, compute the forces on atoms in those configurations using VASP, and obtain the harmonic force constants for each semiconductor. For the polar compounds InAs and GaAs, we employ density functional perturbation theory to compute a set of Born effective charges and the high-frequency dielectric tensor to account for Coulombic interactions \cite{wang_mixed-space_2010}. Those ingredients allow us to obtain the compound's phonon spectrum. A larger set of supercell calculations is used to compute the relevant third-order derivatives of the potential energy with respect to atomic displacements. Finally, all elements are combined to obtain a relaxation time for each phonon mode, including both phonon-phonon processes and isotopic scattering. The last two steps are performed using open-source software developed by some of us and documented in full detail elsewhere \cite{ShengBTE_2014}. For the supercells, $\Gamma$-point-only DFT runs are adequate. We include neighbors up to the fifth coordination shell in our third-order calculations and use a $32\times 32\times 32$ wavevector grid, which yields fully converged values of the thermal conductivity. Our method requires no experimental input and is fully parameter free. The first-principle lattice constants, phonon spectra and room-temperature thermal conductivities agree well with values from the literature.
\par
For a disordered binary alloy A$_{x}$B$_{1-x}$ we operate under the virtual crystal approximation, which has been succesfully tested in similar settings \cite{MgSiSn,katcho_lattice_2012}. Lattice constants, second- and third-order interatomic force constants and dielectric parameters are taken as weighted averages of their values for A and B, with weights $x$ and $1-x$. Mass disorder in the alloy is treated in the same way as isotopic disorder in pure compounds \cite{tamura_isotope_1983}, as described in Ref. \onlinecite{ShengBTE_2014}.
\par
The materials considered in this paper are Si, In$_{0.53}$Ga$_{0.47}$As and Si$_{0.82}$Ge$_{0.18}$. The calculated bulk thermal parameters are listed in Table \ref{tabmaterials}.\npar%
\begin{table}[!htb]
\caption{Thermal properties obtained from ab initio phonon calculations.}\label{tabmaterials}
\vspace{2mm}
\begin{tabular}{ccccc}
\hline
 & $\kappa$ & $C$ & $D = \kappa/C$ & $\bar{v}$ (Eq. \ref{vbar}) \\
Material & [W/m-K] &[MJ/m$^3$-K] & [mm$^2$/s] & [m/s] \\
\hline
Si & 166 & 1.62 & 103 & 1578\\
In$_{0.53}$Ga$_{0.47}$As & 8.56 & 1.56 & 5.49 & 429\\
Si$_{0.82}$Ge$_{0.18}$ & 10.7 & 1.66 & 6.46 & 854\\
\hline
\end{tabular}
\end{table}
\subsection*{Results}
First, we investigate the transient evolution of the thermal energy MSD. The ab initio calculations provide 196,608 phonon modes (six branches over a $32 \times 32 \times 32$ wavevector grid) whose properties are inserted into (\ref{variancemaster}) to calculate the MSD. Time domain curves, obtained by transforming $\sigma^2(s)$ with a standard Gaver-Stehfest Laplace inversion scheme \cite{GS1,GS2}, are plotted in Fig. \ref{1406_7341_v2_fig1}.
\myfig[!htb]{width=0.47\textwidth}{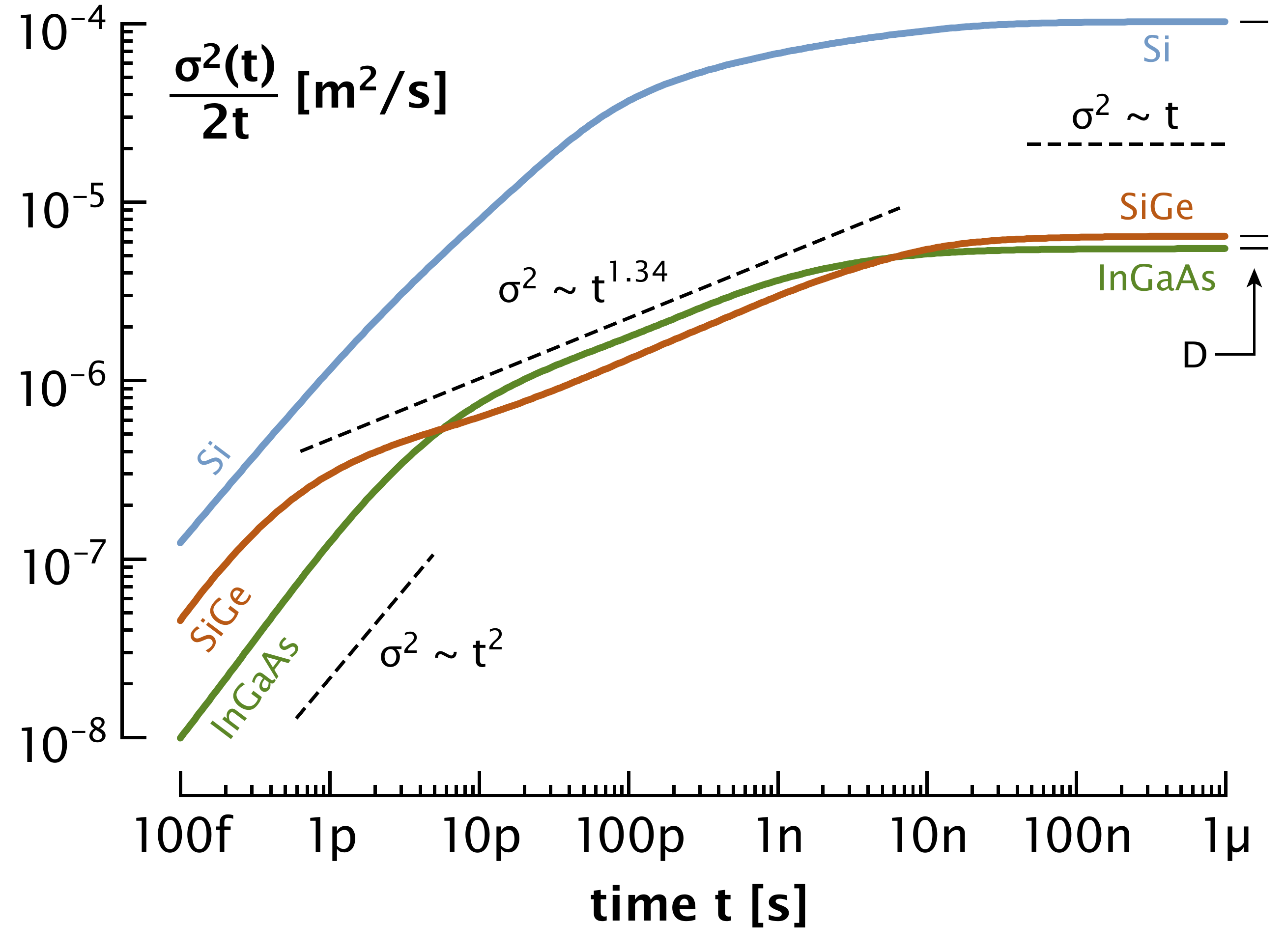}{Renormalised mean square displacement of thermal energy obtained from BTE solution Eq. (\ref{variancemaster}) with ab initio phonon dispersions and scattering rates at room temperature. The emergence of a superdiffusive regime with time exponent $\beta \simeq 1.34$ is clearly apparent for the alloy materials.}
\npar%
Intuitively, one may expect the MSD time exponent to drop smoothly from 2 to 1 during the transition from purely ballistic ($\sigma^2 \sim t^2$) to purely diffusive ($\sigma^2 \sim t$) transport. Instead, for alloy materials we observe the emergence of a striking regime $\sigma^2 \sim t^{\beta} (1 < \beta < 2)$ where $\beta$ remains virtually stable during several orders of magnitude of time. Transport in which the MSD scales superlinearly with time is typically referred to as superdiffusive \cite{fractalwalks}. Least square fitting of the obtained MSD curves yields $\beta = 1.331$ ($30\,\mathrm{ps} \leq t \leq 700\,\mathrm{ps}$) for InGaAs and $\beta = 1.350$ ($20\,\mathrm{ps} \leq t \leq 2\,\mathrm{ns}$) in SiGe. These time windows \revision{are slightly out of reach of the typical bandwidth of TDTR experiments since the time constant associated with the oscillating heat source, $\tau_{\text{mod}} = (2\pi f_{\text{mod}})^{-1}$, exceeds 8$\,$ns for modulation frequencies $f_{\text{mod}} \leq 20\,$MHz. The measurements, however, do not observe MSD, but are predominantly sensitive to the dynamic response of the semiconductor surface at $f_{\text{mod}}$ \cite{cahillmodel}. We therefore take a closer look at the spatiotemporal dynamics of the energy density to better understand the physics behind quasi-ballistic effects exhibited by alloys in TDTR analysis.}
\npar%
Ideally one wishes to look at the distribution $P(z,t)$ in real space-time domain, but numerical limitations prevent a stable and accurate direct inversion. However, we can identify key dynamics directly from the Fourier-Laplace entity (\ref{Pmaster}). Figure \ref{1406_7341_v2_fig2} shows the magnitude of $|P(\xi,s)|$ versus $|\xi|$ ($P$ is even in $\xi$) at various frequencies $s = j 2\pi f$ for the three considered materials.\npar%
\myfig[!htb]{width=0.49\textwidth}{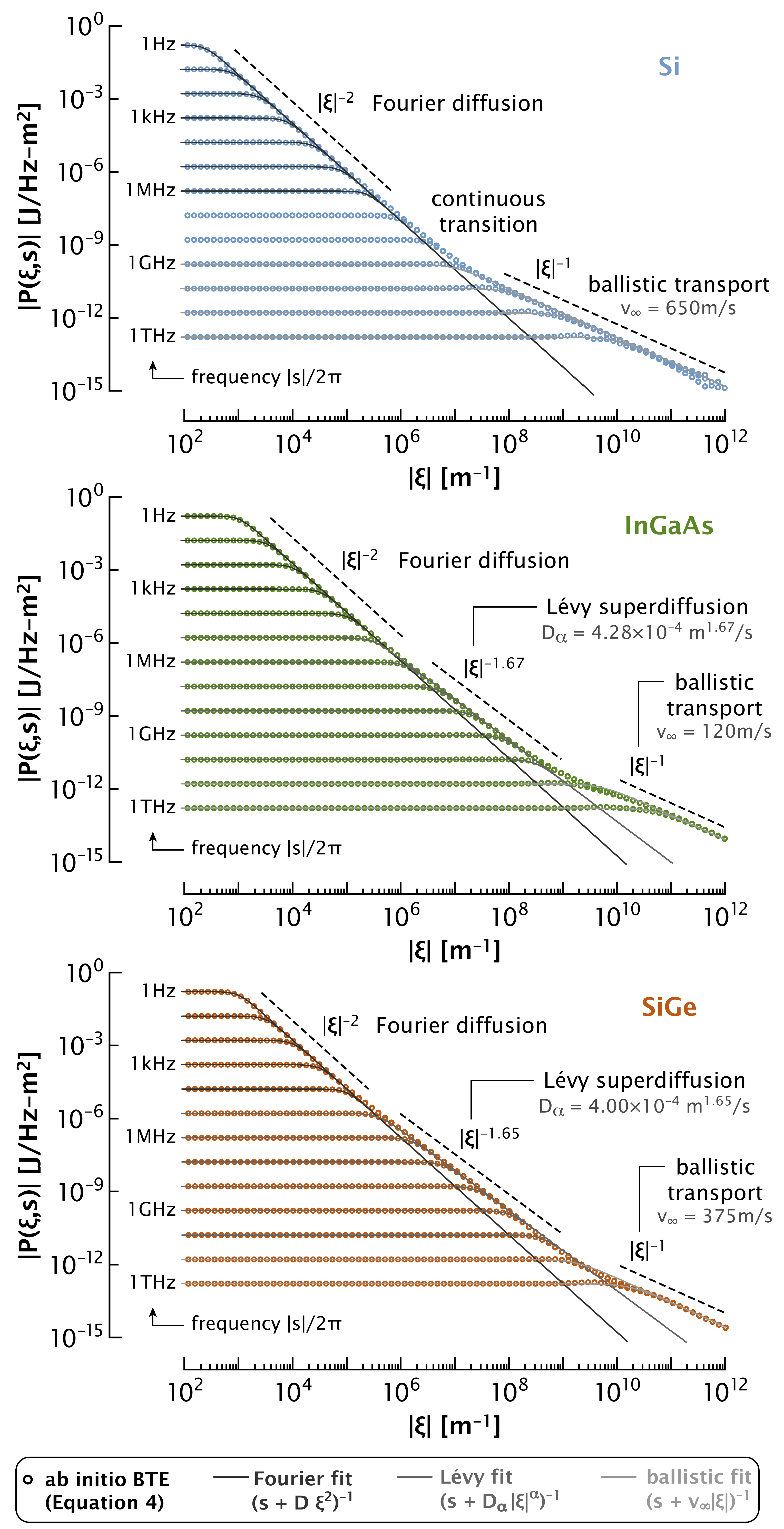}{Magnitude of the energy density distribution in Fourier-Laplace domain, $|P(\xi,s)|$, at various frequencies $s = j 2\pi f$, calculated from (\ref{Pmaster}) with ab initio phonon dispersions and scattering rates. Distinct regimes are visible that each can be described well with a simple analytic expression. The fittings over the intermediate frequency range indicate that quasi-ballistic thermal transport in alloys behaves as L\'evy superdiffusion with fractal dimension $\alpha = 3-\beta$.}
In the diffusive limit, the BTE results recover classical Fourier solutions. Here, $P(z,t)$ is the familiar Gaussian with variance $2Dt$, which in transformed variables reads $P(\xi,s) = 1/(s + D \xi^2)$. In the ballistic limit, the distribution tends to a Lorentzian: $P(\xi,s) \rightarrow 1/(s + v_{\infty}|\xi|)$ hence $P(z,t) \rightarrow v_{\infty}t/[\pi(v_{\infty}^2 t^2 + z^2)]$ \revision{where $v_{\infty}$ can be interpreted as an average collective group velocity along the thermal transport axis. Since the energy transport is jointly governed by a diverse population of phonons, many of which having small group velocities, $v_{\infty}$ is quite a bit lower than the sound velocity}. We note that the actual ballistic distribution will slightly deviate from the asymptotic Lorentz limit since the phonon group velocities contain the energy within a finite $z$ interval that grows with time. At intermediate \revision{spatial and temporal} frequencies in alloys, we find $P(\xi,s) \simeq 1/(s + D_{\alpha} |\xi|^{\alpha}) \Leftrightarrow P(\xi,t) \simeq \exp(-D_{\alpha} t |\xi|^{\alpha})$, where $1 < \alpha < 2$ and $D_{\alpha}$ is a fractional diffusivity constant. These solutions correspond to the characteristic function of a so called L\'evy stable process \cite{levyprocess}, and immediately imply that the quasi-ballistic energy transport is stochastically equivalent to a random walk with fractal dimension $\alpha$ \cite{fractalwalks}. L\'evy-type anomalous diffusion has been encountered across a wide variety of disciplines ranging from travel patterns of foraging animals \cite{animals}, protein movements along DNA chains \cite{DNA}, tracer motion in turbulent fluids \cite{fluidturbulence}, and financial market fluctuations \cite{stockmarket}. L\'evy processes in which finite transition velocities are enforced, as would be appropriate in the context of phonon dynamics, are known to induce an MSD $\sigma^2(t) \sim t^{3-\alpha}$ \cite{levyvariance}. This suggests the superdiffusive exponent and fractal dimension are directly interrelated as $\alpha = 3-\beta$. Figure \ref{1406_7341_v2_fig2} confirms that for InGaAs and SiGe, whose MSDs exhibit $\beta \simeq 1.34$, the quasi-ballistic BTE energy density can indeed be fitted accurately using $\alpha \simeq 1.66$. Contrary to the Gaussian energy densities associated with Fourier diffusion, L\'evy distributions have `fat tails' that spatially decay as a power law $P(z \rightarrow \infty,t) \sim |z|^{-(1+\alpha)}$. The response at the heat source takes the form $P_0(t) = P(z=0,t) \sim t^{-1/\alpha}$, \revision{as in pure L\'evy regime we have}
\begin{equation}
P_{0,L}(s) = [\alpha \sin(\pi/\alpha) D_{\alpha}^{1/\alpha} s^{1 - 1/\alpha}]^{-1} \label{PsurfLevy}
\end{equation}
\revision{which decays less steeply with frequency compared to the Fourier solution}
\begin{equation}
P_{0,F}(s) = [2 \sqrt{sD}]^{-1} \label{PsurfFourier}
\end{equation}
The BTE solutions in alloys at the heat source, obtained numerically through
\begin{equation}
P_0(s) = \frac{1}{2\pi} \int \limits_{-\infty}^{\infty} P(\xi,s) \mathrm{d}\xi \quad , \label{Psurface}
\end{equation}
are indeed found to exhibit those signature L\'evy characteristics \revision{(Fig. \ref{1406_7341_v2_fig3})}.
\myfig[!htb]{width=0.47\textwidth}{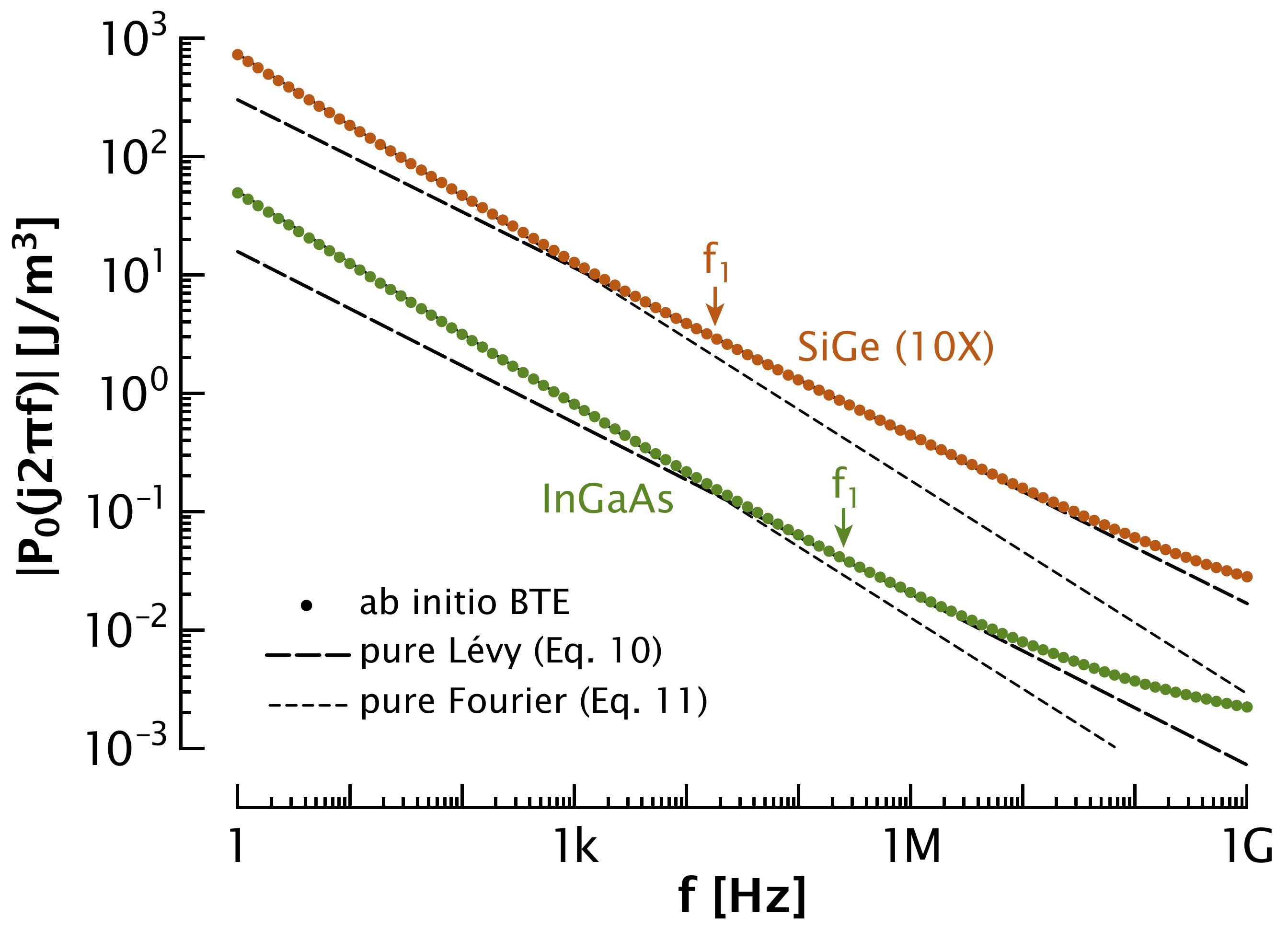}{\revision{Frequency domain thermal response at the heat source in semiconductor alloys. SiGe results are upscaled by a factor 10 for visual separation of the curves. The arrows indicate the predicted threshold frequency $f_1$, calculated from Eq. (\ref{levywindowf}), for onset of pure L\'evy dynamics.}}
\par
\revision{The energy density can be approximated quite well by the form $P(\xi,s) \simeq [s + \psi(\xi)]^{-1}$ across all three distinct transport regimes. Observing that $\psi(\xi) = 1/P(\xi,s \rightarrow 0)$, we obtain from (\ref{Pmaster})
}
\begin{equation}
\psi(\xi) = \xi^2 \cdot \frac{\sum \limits_{k_z \geq 0} \frac{C_k \Lambda_{\shortparallel,k}^2}{\tau_k [1 + \xi^2 \Lambda_{\shortparallel,k}^2]}}{\sum \limits_{k_z \geq 0} \frac{C_k}{1 + \xi^2 \Lambda_{\shortparallel,k}^2}} \label{psixi}
\end{equation}
\revision{This function, which governs the asymptotic `backbone' of the curves in Fig. \ref{1406_7341_v2_fig2}, is helpful to graphically identify $D, \alpha, D_{\alpha}$ and $v_{\infty}$ (Fig. \ref{1406_7341_v2_fig4}).}
\myfig[!htb]{width=0.47\textwidth}{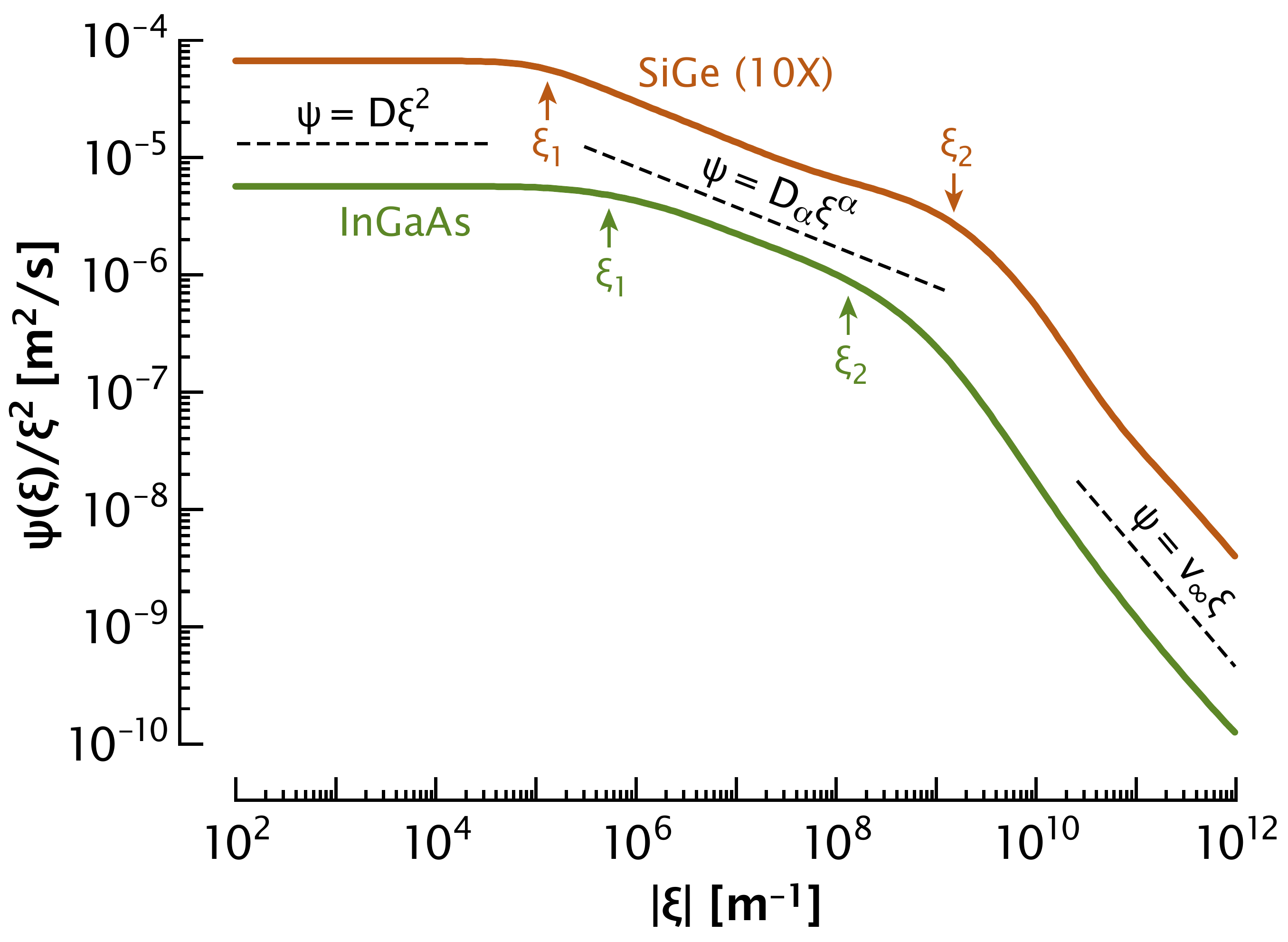}{\revision{The function $\psi(\xi)$, defined by Eq. (\ref{psixi}) and which governs the asymptotic backbone of the energy density curves in Fig. \ref{1406_7341_v2_fig2}, offers convenient identification of the three distinct transport regimes in semiconductor alloys. The SiGe result is upscaled by a factor 10 for visual separation of the curves. The arrows demarkate the L\'evy window $[\xi_1,\xi_2]$ as estimated from Eq. \ref{levywindowxi}.}}
\par
\revision{The spatial frequency range $[\xi_1,\xi_2]$ that exhibits L\'evy dynamics can be easily estimated from the cross-over points between neighbouring regimes:
}
\begin{eqnarray}
D \xi_1^2 = D_{\alpha} \xi_1^{\alpha} & \quad \rightarrow \quad & \xi_1 = (D_{\alpha}/D)^{\frac{1}{2-\alpha}} \nonumber \\
D_{\alpha} \xi_2^{\alpha} = v_{\infty} \xi_2 & \quad \rightarrow \quad & \xi_2 = (v_{\infty}/D_{\alpha})^{\frac{1}{\alpha - 1}} \label{levywindowxi}
\end{eqnarray}
\revision{In turn, this directly determines the frequency bandwidth $[f_1,f_2]$ over which L\'evy effects are observable, since the energy density curve starts tying into the backbone at cross-over point $\xi_c$ fulfilling $|s| = \psi(\xi_c)$. We find}
\begin{eqnarray}
f_1 & = & \frac{D \xi_1^2}{2\pi} = \frac{D}{2\pi} \cdot \left( \frac{D_{\alpha}}{D} \right)^{\frac{2}{2-\alpha}} \nonumber \\
f_2 & = & \frac{v_{\infty} \xi_2}{2\pi} = \frac{v_{\infty}}{2 \pi} \cdot \left(\frac{v_{\infty}}{D_{\alpha}}\right)^{\frac{1}{\alpha - 1}} \label{levywindowf}
\end{eqnarray}
\revision{Time domain thermoreflectance measurements are primarily sensitive to the thermal response of the semiconductor surface at the modulation frequency \cite{cahillmodel}. Recalling (\ref{Psurface}) and the steep algebraic decay of $P(\xi,s)$, we see that experiments at modulation frequencies near the lower threshold $f_1 \leq f_{\text{mod}} \ll f_2$ will be dominated by pure L\'evy dynamics. The ab initio results for $D$, $\alpha$, $D_{\alpha}$ and $v_{\infty}$ determined above produce $f_1$ values of 255$\,$kHz for InGaAs and 18$\,$kHz in SiGe, while $f_2$ is in the GHz range for both cases. The earlier presented Fig. \ref{1406_7341_v2_fig3} confirms that the source response matches the pure L\'evy solution at frequencies near the calculated threshold values. In fact, $f_1$ provides a conservative estimate: the source response already approaches L\'evy dynamics at frequencies that are roughly an order of magnitude smaller. This can easily be understood from the fact that the intersection of the dashed lines, governed by the crossover of Eqs. (\ref{PsurfLevy}) and (\ref{PsurfFourier}), occurs at $f_0 = [(\alpha/2)\sin(\pi/\alpha)]^{\frac{2\alpha}{2-\alpha}} f_1 \simeq f_1/10$. We further note that with $\alpha \simeq 5/3$, the threshold frequency varies with the sixth power of the $D_{\alpha}/D$ ratio, and can thus be expected to be quite sensitive to the crystal quality and purity of actual material samples. In addition, these threshold frequencies obtained from 1D modelling should not be regarded as sharp cutoffs but rather as approximate estimates for actual measurements, as we can expect that 3D heat flow will slightly shift the cross-over between regimes. L\'evy identification of our TR experiments spanning the 1--20$\,$MHz bandwidth (briefly summarised in a later section and described in full detail in part II \cite{part2}) indicates $f_1$ values of 920$\,$kHz and 1.46$\,$MHz in our InGaAs and SiGe samples respectively. The measurements therefore predominantly probe the pure L\'evy regime, enabling reliable experimental extraction of $\alpha$ and $D_{\alpha}$.
}
\subsection*{Physical origin of L\'evy dynamics}
The question still remains why alloys exhibit a superdiffusive \revision{L\'evy} regime, and what physically determines the associated MSD time exponent $\beta$ \revision{and fractal space dimension $\alpha$}. Here, we show that the quasi-ballistic dynamics of a material originate in its dominant phonon scattering mechanism. To simplify the analysis, we limit ourselves to a single phonon branch with constant group velocity $v_0$ in an ideal isotropic crystal with spherical BZ. We should expect this approach to capture the essential trends occurring in actual media, as a major fraction of thermal conduction is governed by acoustic phonons on quasi straight branch segments near the zone center. At room temperature, energies of the dominant acoustic phonons typically do not exceed $k_B T$. Under these circumstances, $\hbar \omega \frac{\partial f_{\text{BE}}}{\partial T}$ varies by less than 8\% across the modes, and we can simply assume a constant mode capacity ($C_k \equiv C_0$) with good approximation. We consider a single dominant phonon scattering mechanism of the form:
\begin{equation}
\tau \sim \omega^{-n} \Rightarrow \tau_{k} = \tau_{\text{min}} \left( \frac{ka}{\pi} \right)^{-n} \label{singlebranchtau1}
\end{equation}
in which $n$ is order of the mechanism (not necessarily integer), $a$ is the lattice constant and $k = ||\vec{k}||$ the wavevector norm $0 \leq k \leq \frac{\pi}{a}$. First principle calculations in Si \cite{firstprincipleSi} have suggested that Umklapp processes, which dominate its bulk thermal conductivity at room temperature, are characterised by $n = 3$, although we note that other works have also inferred $n=2$ \cite{holland} and $n=4$ \cite{broido} relations. Mass impurity (Rayleigh) scattering, dominating alloy behaviour, ideally obeys $\tau \sim v_{\omega}^3 \cdot \omega^{-4}$ which corresponds to $n = 4$ under the assumed linear dispersion. Inserting the single branch relations into the BTE variance (\ref{variancemaster}) and performing BZ volume integration in spherical coordinates yields
\begin{equation}
\sigma^2(s) = \frac{2}{3} v_0^2 \tau_{\text{min}} \cdot \frac{\int \limits_{0}^{1} \frac{\tilde{k}^{n+2} \mathrm{d} \tilde{k}}{(\tilde{k}^{n} + s\tau_{\text{min}})^2}}{s^2 \int \limits_{0}^{1} \frac{\tilde{k}^{n+2} \mathrm{d} \tilde{k}}{\tilde{k}^{n} + s\tau_{\text{min}}}}
\end{equation}
where $\tilde{k} = ak/\pi$ is a dimensionless wavevector norm. For $s \tau_{\text{min}} \ll 1$, i.e. once a sufficient fraction of the phonon population have undergone scattering events to break up the purely ballistic regime, one finds $\sigma^2(s) \sim s^{-3+3/n}$ when $n>3$. The time domain counterpart is a power law $\sigma^2(t) \sim t^{\beta_n}$ with superdiffusive exponent
\begin{equation}
\beta_{n} = 2 - \frac{3}{n} \quad , \quad n > 3 \label{beta_n}
\end{equation}
We point out that simple scattering relations of the form (\ref{singlebranchtau1}) produce arbitrarily large relaxation times near the zone center. Bulk thermal conductivity of the single branch model diverges for $n > 2$, and the regular diffusive transport regime is never recovered. In reality, scattering times are physically bounded to a finite range $\tau_{\text{min}} \leq \tau \leq \tau_{\text{max}}$ with $\tau_{\text{max}} \gg \tau_{\text{min}}$. As a result, a superdiffusive regime with $\beta \simeq \beta_{n}$ is only maintained over a finite time window in the actual medium, as observed earlier in Fig. \ref{1406_7341_v2_fig1}.
\npar%
We can additionally link $\beta$ and the temporal extent of the superdiffusive window \revision{in the MSD} to the shape of the cumulative conductivity function $\kappa_{\Sigma}(\tau) = \sum \limits_{\tau_k \leq \tau} \kappa_k$. A single branch model with scattering relation (\ref{singlebranchtau1}) provides
\begin{equation}
\kappa_{\Sigma}(\tau) \sim \int \limits_{(\tau_{\text{min}}/\tau)^{1/n}}^{1} \frac{\mathrm{d}\tilde{k}}{\tilde{k}^{n-2}} \label{kcumultau}
\end{equation}
from which we obtain
\begin{eqnarray}
n = 2 & : & \kappa_{\Sigma} \sim 1 - \sqrt{\frac{\tau_{\text{min}}}{\tau}} \nonumber \\
n = 3 & : & \kappa_{\Sigma} \sim \ln \left( \frac{\tau}{\tau_{\text{min}}}\right) \nonumber \\
n > 3 \text{ (incl. Rayleigh)} & : & \kappa_{\Sigma} \sim \left( \frac{\tau}{\tau_{\text{min}}} \right)^{\beta_n - 1} - 1 \label{kcumulmodel}
\end{eqnarray}
where we used (\ref{beta_n}) for the last case.
\par
\revision{The single branch model also helps reveal the physical origin of the L\'evy window in the energy density. Under the assumptions outlined earlier, we have $\Lambda_{\shortparallel,k} = v_0 \tau_k \cos \theta = \Lambda_{\text{min}} \tilde{k}^{-n} \cos \theta$. Here, we will also explicitly enforce a physically bounded MFP $\Lambda \leq \Lambda_{\text{max}}$ by excluding an appropriately chosen small sphere around the zone center. This is easily achieved by introducing a lower cutoff wavevector norm $\tilde{k}_{\text{min}} = (\Lambda_{\text{min}}/\Lambda_{\text{max}})^{1/n}$ in the BZ integration. The Debye branch equivalent of (\ref{psixi}) reduces to}
\begin{eqnarray}
\psi(\xi) & = & \frac{\int \limits_{\tilde{k}_{\text{min}}}^{1} \mathrm{d}\tilde{k} \int \limits_{0}^{\pi/2} \frac{\tilde{k}^{n+2} \xi^2 \Lambda_{\text{min}}^2 \cos^2 \theta \sin \theta \mathrm{d} \theta}{\tau_{\text{min}} (\tilde{k}^{2n} + \xi^2 \Lambda_{\text{min}}^2 \cos^2 \theta)}}{\int \limits_{\tilde{k}_{\text{min}}}^{1} \mathrm{d}\tilde{k} \int \limits_{0}^{\pi/2} \frac{\tilde{k}^{2n+2} \sin \theta \mathrm{d} \theta}{\tilde{k}^{2n} + \xi^2 \Lambda_{\text{min}}^2 \cos^2 \theta}} \nonumber \\
& = & \frac{\tilde{\xi} \int \limits_{\tilde{k}_{\text{min}}}^{1} \tilde{k}^{n+2} \left[ 1 - (\tilde{k}^n/\tilde{\xi}) \arctan (\tilde{\xi}/\tilde{k}^n) \right] \mathrm{d}\tilde{k}}{\tau_{\text{min}} \int \limits_{\tilde{k}_{\text{min}}}^{1} \tilde{k}^{n+2} \arctan (\tilde{\xi}/\tilde{k}^n) \mathrm{d}\tilde{k}} \label{psisinglebranch}
\end{eqnarray}
\revision{with $\tilde{\xi} = \Lambda_{\text{min}} \, \xi$ a dimensionless spatial frequency. When $\xi \ll \Lambda_{\text{max}}^{-1}$, we have $\tilde{\xi} \ll \tilde{k}^n$ over the entire integration domain. This leads to $\psi(\xi) \sim \xi^2$ which signifies a regular Fourier diffusion regime. Conversely, when $\xi \gg \Lambda_{\text{min}}^{-1}$ such that $\tilde{\xi} \gg \tilde{k}^n$ over the entire integration domain, $\psi(\xi) \simeq 2 v_0 \xi / \pi$ which signifies a ballistic regime with $v_{\infty} = (2/\pi) v_0$. In the intermediate range $\Lambda_{\text{max}}^{-1} \leq \xi \leq \Lambda_{\text{min}}^{-1}$, the arctangents sweep across their entire range over the integration domain. The resulting failure of series expansions prevents us from deriving a simple closed form expression for $D_{\alpha}$, but numerical evaluations of (\ref{psisinglebranch}) for various $n$ values reveal a quasi-ballistic L\'evy regime with fractal dimension $\alpha_n \simeq 1 + 3/n$, in accordance with the expected relation $\alpha_n + \beta_n = 3$. Combining this result with the trends for the MFP counterpart of (\ref{kcumultau}), $\kappa_{\Sigma}(\Lambda)$, also suggests that the earlier determined L\'evy window $[\xi_1,\xi_2]$ in the energy density is associated to an MFP range $[\xi_2^{-1},\xi_1^{-1}]$ over which the cumulative conductivity evolves as $\kappa_{\Sigma}(\Lambda) \sim \Lambda^{\beta_n-1}$.}
\par
The interconnections between the scattering relation $\tau(\omega)$, the cumulative conductivity $\kappa_{\Sigma}$ \revision{resolved for phonon scattering times and MFPs} and the \revision{characterisitic L\'evy} exponents $\alpha$ and $\beta$ we have derived for a single phonon branch with linear dispersion are preserved quite well in realistic media (Fig. \ref{1406_7341_v2_fig5}).
\npar%
\myfigwide[!htb]{width=\textwidth}{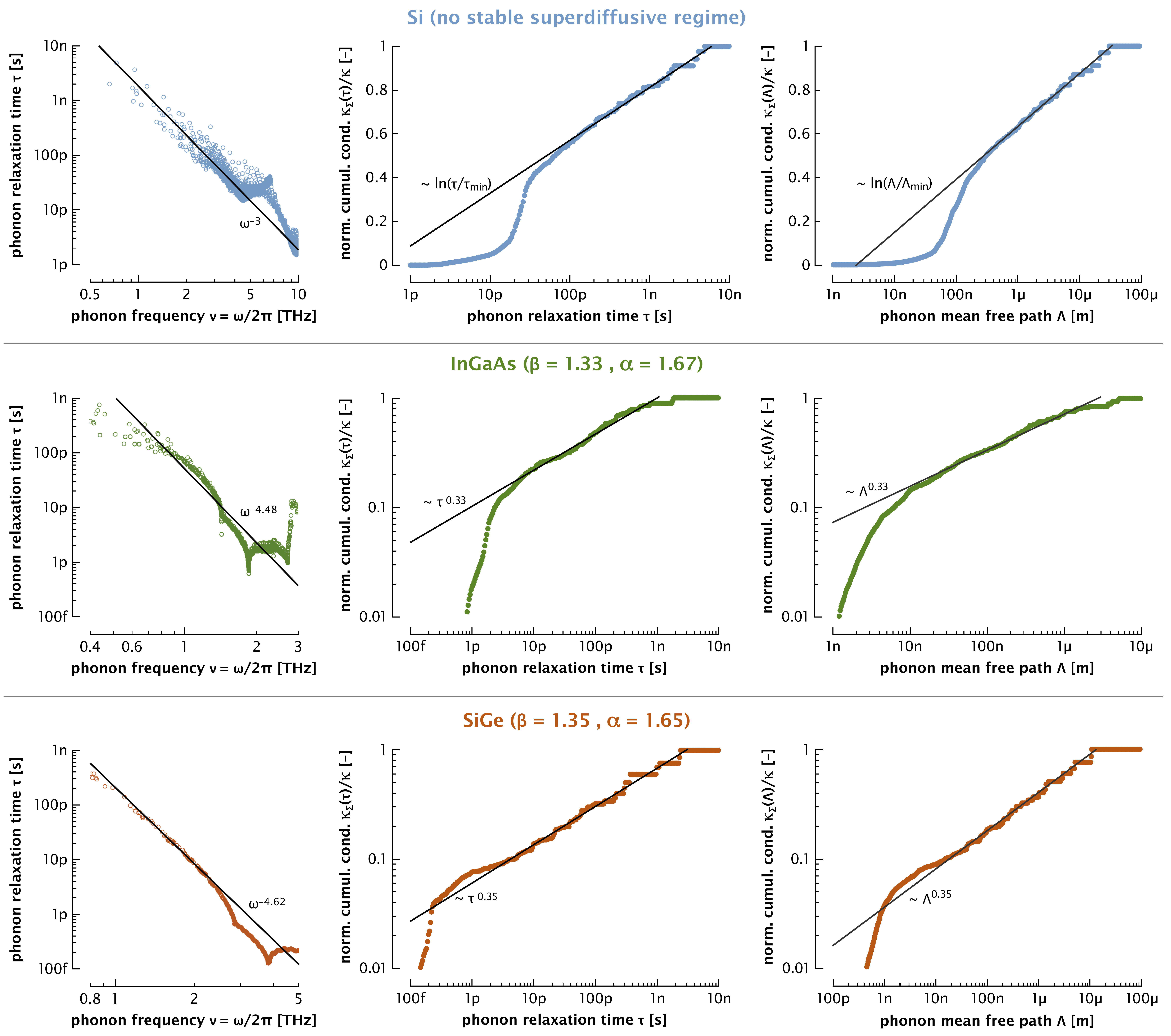}{Interconnection between dominant scattering order $\tau \sim \omega^{-n}$, cumulative conductivity $\kappa_{\Sigma}(\tau,\Lambda) \sim (\tau,\Lambda)^{\gamma}$ and characteristic L\'evy dynamics $\sigma^2(t) \sim t^{\beta} \, , \, \psi(\xi) \sim \xi^{\alpha}$. Ab initio results (circles) agree well with single branch model predictions (lines) $\beta = 2 - \frac{3}{n} \leftrightarrow n = \frac{3}{2-\beta}$ and $\gamma = \beta - 1 = 2-\alpha$.}
For Si, we find that our ab initio scattering rates can be fitted quite well by a scattering relation $\tau \sim \omega^{-3}$. The upper portions of the cumulative conductivity curve show a clear logarithmic dependence on scattering time \revision{and MFP}, as predicted by the ideal $n=3$ case. Initial parts of the $\kappa_{\Sigma}$ curves are governed by higher energy modes with small velocity, and therefore do not obey the simple linear dispersion model, as could be expected. As far as alloys are concerned, single branch expressions (\ref{beta_n}) and (\ref{kcumulmodel}) suggest that the SiGe superdiffusion exponent $\beta = 1.35$ theoretically corresponds to $n = 4.61$ and $\kappa_{\Sigma} \sim \tau^{0.35} \, , \, \Lambda^{0.35}$. Both relations provide good fits to the actual phonon data as shown in Fig. \ref{1406_7341_v2_fig5}. Similar observations hold for InGaAs. It is noteworthy that the relaxation time range over which $\kappa_{\Sigma}(\tau)$ closely follows a power law has a near one-to-one correspondence to the superdiffusive time window in the MSD observed earlier in Fig. \ref{1406_7341_v2_fig1}. \revision{In analog fashion, the spatial window $[\xi_2^{-1},\xi_1^{-1}]$ calculated from (\ref{levywindowxi}), amounting to [7$\,$nm, 2$\,\mu$m] in InGaAs and [0.6$\,$nm, 8$\,\mu$m] in SiGe, matches near perfecly with the MFP range over which $\kappa_{\Sigma}(\Lambda)$ obeys a power law}. As a summarising rule of thumb, we can say that thermal transport in a material whose cumulative conductivity curves $\kappa_{\Sigma}$ have a stable slope $\gamma$ in double logarithmic scale for $\tau_1 \leq \tau \leq \tau_2$ \revision{and $\Lambda_1 \leq \Lambda \leq \Lambda_2$} will exhibit a superdiffuse window $\tau_1 \lesssim t \lesssim \tau_2$ in the MSD with exponent $\beta = \gamma + 1$  \revision{and L\'evy window $\Lambda_2^{-1} \lesssim \xi \lesssim \Lambda_1^{-1}$ in the energy density with fractal dimension $\alpha = 2 - \gamma$}.
\subsection*{Implications for transient laser thermoreflectometry}
Regular diffusive transport is well known to correspond to Brownian motion \cite{brownian}, a stochastic process with fractal dimension 2. Many observations of quasi-ballistic heat flow are interpreted using so called modified Fourier theory \cite{cahill,siemens,minnich,malen}, which explains the anomalous heat conduction phenomenologically in terms of a reduced effective thermal conductivity. Such an approach still inherently maintains the assumption of purely diffusive transport dynamics. The results we have presented here, however, demonstrate that the quasi-ballistic regime in alloys is characterised by superdiffusive L\'evy motion with fractal dimension $\alpha < 2$. The associated energy density distributions, $P(\xi,s) \simeq 1/(s + D_{\alpha} |\xi|^{\alpha})$, are decidedly non-Gaussian in space domain, and cannot be described by a modified Fourier solution $P(\xi,s) = 1/(s + D_{\text{eff}} \, \xi^2)$ due to fundamental mismatch of the $\xi$ exponent. The resulting shortcomings of modified Fourier analyses of thermoreflectance experiments are illustrated in detail in part II of the paper \cite{part2}.
\par%
Hua and Minnich have recently shown that transient thermal grating (TTG) experiments probe a so called weakly quasi-ballistic regime in which the use of a modified Fourier approach is formally justified by the BTE \cite{minnichBTE}. As we have shown above, this is clearly not the case for time/frequency domain thermoreflectance (TR) measurements. The different behaviour can be attributed to important distinctions in experimental configuration and boundary conditions. TTG studies the transient decay of a spatially periodic temperature input directly at the semiconductor surface. The thermal gradients are predominantly in-plane and stretch across the grating wavelength $\lambda$, which typically measures several hundreds of nanometers or more. To a first order, this measurement configuration essentially probes the energy density distribution at a single spatial Fourier variable $\xi_{\lambda} = 2 \pi / \lambda$. As a result, the recorded thermal \revision{response decays exponentially in time and is therefore physically indistinguishable from regular Fourier diffusion dynamics with adjusted diffusivity $D_{\text{eff}} = \psi(\xi_{\lambda})/\xi_{\lambda}^2$}. TR experiments, on the other hand, capture the response to temporally periodic energy impulses. For typical laser spot sizes, the predominant thermal gradient occurs cross-plane over the thermal penetration depth inside the semiconductor. In a crucial difference with the TTG configuration, the thermal field is not spatially periodic in the dominant thermal transport direction. As a result, the semiconductor surface response under cross-plane heat flow is governed by a wide spectrum of $\xi$ values, as symbolised formally by Eq. (\ref{Psurface}). \revision{The quasi-ballistic single pulse response decays as $P_0 \sim t^{-1/\alpha}$ as mentioned earlier. Contrary to TTG, this time signature is inherently different from the Fourier diffusion counterpart $P_0 \sim t^{-1/2}$, even at time scales far exceeding typical phonon relaxation times. The thermal transients observed by TR experiments preserve characteristic L\'evy dynamics, enabling this technique to actively measure the fractal dimension and fractional diffusivity of the quasi-ballisitic transport.}
\subsection*{Experimental validation}
As mentioned in the Introduction, modified Fourier interpretation of TR measurements on semiconductor alloys produces effective conductivities that drop significantly with increasing laser modulation frequency \cite{cahill}. Interestingly, in retrospect this behaviour is a direct manifestation of fractal L\'evy transport. The essence of the connection can be easily understood in terms of the dynamics at the semiconductor surface. \revision{As demonstrated earlier, the response probed by the typical measurement bandwidth is dominated by the pure L\'evy regime. Interpreting the associated semiconductor surface response (\ref{PsurfLevy})} with a modified Fourier solution $P_0(s) = [2 \sqrt{D_{\text{eff}} s}]^{-1}$ suggests that $D_{\text{eff}}(s) \sim s^{1-2/\alpha}$, and more specifically
\begin{equation}
\kappa_{\text{eff}} = 2 \pi C \cdot \left( \frac{\alpha}{2} \right)^2 \cdot \sin^2 \left( \frac{\pi}{\alpha} \right) \cdot \left( \frac{D_{\alpha}}{2 \pi}\right)^{{2}/{\alpha}} \cdot f_{\text{mod}}^{1 - 2/\alpha} \label{keffpowerlaw}
\end{equation}
Based on the BTE results obtained above, we would therefore expect the effective conductivity in InGaAs and SiGe to drop by roughly 40\% over the 1--10$\,$MHz modulation range ($\kappa_{\text{eff}}(10\,\mathrm{MHz})/\kappa_{\text{eff}}(1\,\mathrm{MHz}) = 10^{1-2/1.66} \approx 0.62$), which is quite similar to the actual reduction observed experimentally. A closer look at our own TDTR measurements shows that the effective conductivity of semiconductor alloys can be fitted quite accurately by a power law (Fig. \ref{1406_7341_v2_fig6}). \revision{We note that the presented experimental values result from analysing the measurement data with a 3D Fourier model that accounts for the Gaussian shape of pump and probe laser beams, heat spreading through the metal transducer, and thermal contact resistance of the metal/semiconductor interface.} \npar%
\myfig[!htb]{width=0.4\textwidth}{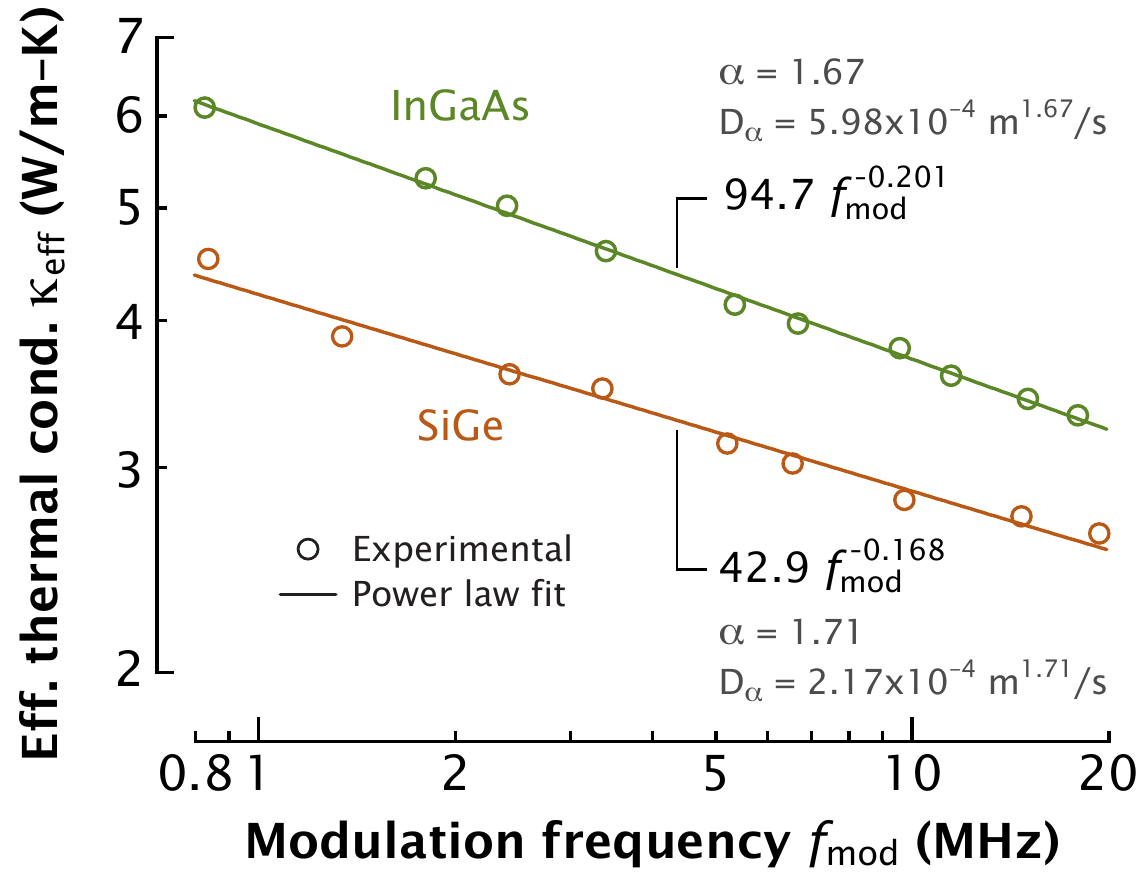}{The frequency dependence of effective thermal conductivity, observed in time domain thermoreflectance experiments on semiconductor alloys, is a direct manifestation of fractal L\'evy transport. A power law fit, suggested by basic 1D model prediction (\ref{keffpowerlaw}), provides accurate estimates of the key metrics of the quasi-ballistic transport.}
Comparing the fits to (\ref{keffpowerlaw}) produces $\alpha = 1.67$ for InGaAs and $\alpha = 1.71$ for SiGe, in good agreement with the ab initio predictions found earlier (1.67 and 1.65). In addition, the extracted fractional diffusivities $D_{\alpha}$ are both within $\pm 45\%$ of the theoretical values from Fig. 2. This too can be considered a very reasonable agreement, given that the measured bulk thermal conductivity deviates from BTE predictions by similar amounts.\npar%
The intricacies of TR experiments are obviously far more involved than simple 1D model expressions. Rigorous analysis of raw measurement data is desirable to determine the fractal dimension directly, and establish a definitive experimental confirmation of superdiffusive L\'evy behaviour. Unfortunately, 1D BTE solutions as derived here to illuminate essential trends are not easily suitable for this purpose. Crystal impurities in real samples cause perturbations in the phonon spectra, while the effects of heat source nonuniformity (Gaussian shaped laser beam) and lateral heat spreading can only be accounted for by 3D modelling. To achieve this, we have developed a phenomenological approach based on truncated L\'evy theory \cite{truncation2}. Our method captures the essential physics of the BTE solutions yet offers sufficient flexibility for full 3D analysis of experimental data. The methodology and performance of the formalism are presented in full detail in part II of the paper \cite{part2}. Here, we just mention that our truncated L\'evy model provides accurate fits to raw measurement data across the entire 1--20$\,$MHz modulation range without requiring any frequency dependent `effective' thermal parameters. This identification produced $\alpha=1.67$ for InGaAs and $\alpha = 1.69$ for SiGe, once again in close agreement with previously derived values.
\subsection*{Conclusions}
In summary, we investigated the fundamental dynamics of 1D quasi-ballistic heat conduction. Analytical solutions of the BTE with ab initio phonon properties reveal the distinct emergence of L\'evy superdiffusion in semiconductor alloys. Simple algebraic expressions capture the intricate relationships between the superdiffusive time exponent, fractal space dimension, order of the dominant phonon scattering mechanism, and cumulative conductivity functions. Our findings lend fundamental physical support to a novel truncated L\'evy heat formalism we have developed, enabling direct experimental measurements of the L\'evy properties of the quasi-ballistic thermal transport in alloys.
\subsection*{Acknowledgements}
B.V. thanks Jesse Maassen at the Birck Nanotechnology Center for helpful discussions. B.V. and A.S. acknowledge funding from the Center for Energy Efficient Materials, an Energy Frontier Research Center funded by the U.S. Department of Energy, Office of Basic Energy Sciences under Award Number DE-SC0001009.


%

\end{document}